\documentclass{article}

\usepackage[final,nonatbib]{neurips_2019}
\usepackage{amssymb,amsmath,enumerate,url,fullpage}
\usepackage{verbatim,graphicx,balance}
\usepackage{siunitx} 
\usepackage{comment,appendix}
\usepackage{mathtools}
\usepackage{algorithmic}
\usepackage{algorithm2e} 
\usepackage{xcolor}
\DeclarePairedDelimiter{\ceil}{\lceil}{\rceil}
\newtheorem{definition}{Definition}

\usepackage[utf8]{inputenc} 
\usepackage[T1]{fontenc}    
\usepackage{hyperref}       
\usepackage{url}            
\usepackage{booktabs}       
\usepackage{amsfonts}       
\usepackage{nicefrac}       
\usepackage{microtype}      
\usepackage{makecell}

\title{HCA-DBSCAN: HyperCube Accelerated Density Based Spatial Clustering for Applications with Noise}

\author{%
    Vinayak Mathur 
   \\
  College of Information and Computer Sciences\\
  University of Massachusetts, Amherst\\
  \texttt{vinayak@cs.umass.edu} \\
   \AND
     Jinesh Mehta \\ 
   Department of Information and Communication Technology \\
   Manipal Institute of Technology \\
   \texttt{mehtajineshs@gmail.com} \\
   \AND
   Sanjay Singh \\
   Department of Information and Communication Technology \\
   Manipal Institute of Technology \\
   \texttt{sanjay.singh@manipal.edu} \\
}

\begin{document}

\maketitle

\begin{abstract}
  Density-based clustering has found numerous applications across various domains. The Density-Based Spatial Clustering of Applications with Noise (DBSCAN) algorithm is capable of finding clusters of varied shapes that are not linearly separable, at the same time it is not sensitive to outliers in the data. Combined with the fact that the number of clusters in the data are not required apriori makes DBSCAN really powerful. Slower performance ( $O(n^2)$) limits its applications. In this work we present a new clustering algorithm, the HyperCube Accelerated DBSCAN(HCA-DBSCAN) which uses a combination of distance based aggregation by overlaying the data with customized grids. We use representative points to reduce the number of comparisons that need to be computed. Experimental results show that the proposed algorithm achieves a significant run time speedup of up to 58.27\% when compared to other improvements that try to reduce the time complexity of the DBSCAN algorithm. 
  
\end{abstract}

\section{Introduction}

Clustering algorithms group data based on a similarity index, that is the algorithms try to increase the intra-cluster similarity, and at the same time reduce the inter-cluster similarity. This class of unsupervised learning algorithms have found applications both within and outside the machine learning community. Clustering algorithms are broadly classified into three categories - Partitioning, Hierarchical, and Density based. In this work we explore density based algorithms where a cluster is defined as a region in which the density of data objects exceeds some threshold. Density-Based Spatial Clustering of Applications with Noise (DBSCAN) was one of the first density-based algorithm proposed in 1996 by Ester et al. \cite{dbscan}. Due to its importance in both theory and applications, this algorithm was awarded the Test of Time Award at SIGKDD 2014 \cite{award}. One of the chief reason that  DBSCAN algorithm has been so popular is that unlike the partitioning approach the number of clusters need not be specified by the user, this is really useful in scenarios where much information about the data is not available apriori. The algorithm is capable of finding clusters of arbitrary shape in the data set which may not be linearly separable at the same time it is not limited to clusters of a specific shape (for example spherical clusters in the partitioning approach). Additionally, the algorithm is not sensitive to outliers. These three properties combined give DBSCAN a significant edge over some of the other clustering algorithms. DBSCAN scales well with higher dimension data and runs with a time complexity of $O(n^2)$; however, if spatial indexing is used this complexity reduces to $O(n\log{n})$. Improving the performance of DBSCAN can open it up to new applications. Different approaches have been tried to achieve this goal, we try to improve performance by reducing the number of comparisons that the algorithm needs to make.

\par
The DBSCAN algorithm visits each point, possibly multiple times. Without the use of spatial indexing, or on degenerated data (e.g., all points within a distance less than $\epsilon$), the worst-case run time complexity remains $O(n^2)$ \cite{wiki:db}. Additionally, spatial indexing methods do not work efficiently for higher dimensional data, where the run time complexity grows from $O(n\log{n})$ to $O(n^2)$ \cite{rtree}. To overcome these issues we propose the HyperCube Accelerated DBSCAN algorithm (HCA-DBSCAN). The HCA-DBSCAN algorithm runs with a temporal complexity of $O(n\log{n})$ (for lower dimensions) and scales to $O(n^{3/2})$ for higher dimensional data. The HCA-DBSCAN algorithm uses a unique combination of a virtual grid, which is imposed on the input data and employs representative points, this significantly reduces the number of comparisons that need to be made, which translates into a significant run time speedup of up to 58.27\% when compared to other proposed improvements. 
The major contribution of this article is summarized as follows: (i) We propose a variation of DBSCAN that runs with a complexity of $O(n\log{n})$ for lower dimensions and scales to $O(n^{3/2})$ for higher dimensional data. (ii) HCA-DBSCAN maintains 100\% accuracy of the original DBSCAN algorithm while achieving the speedup. (iii) HCA-DBSCAN retains the versatility of the DBSCAN algorithm and can identify clusters with arbitrary shape. (iv) The algorithm achieves a significant speedup in run time when compared to some of the other improvements proposed for the DBSCAN algorithm.


\section{Methodology: HyperCube Accelerated DBSCAN}
\label{sec:c}
In this section we detail how the algorithm performs clustering and achieves the speedup. A detailed flowchart, the formal algorithm, and a complexity analysis are presented as supplementary material.
First, we perform a pre-processing step on the input data and sort the data set in the leading dimension. Ties are broken based on secondary dimensions. This pre-processing speeds up the hypercube allocation as explained below.

\textbf{Merging Condition: }The algorithm checks if the distance between two points is less than the input parameter $\epsilon$. If it is, then the two hypercubes that these two points belong to are merged to belong to the same cluster. \par
\textbf{Overlaying Hypercubes:} A hypercube is a closed, compact, convex figure whose 1-skeleton consists of groups of opposite parallel line segments aligned in each of space's dimensions, perpendicular to each other and of the same length \cite{wiki:hypercube}. So in 1-D, a hypercube is a line segment, a square in 2-D, a cube in 3-D, a tesseract in 4-D and so on. Based on the dimensionality of the data we overlay a virtual hypercube on the points such that the length of the space diagonal of the hypercube is $\epsilon$. This $\epsilon$ corresponds to the $\epsilon$ input parameter of the DBSCAN algorithm which is detailed in the supplementary sections. The key idea for the construction of this grid is that we construct it in such a way that every point in a particular hypercube is guaranteed to belong to the same cluster. This results in a performance improvement as  instead of checking every point against each point in the data set like the original DBSCAN algorithm we need to only check if any one point in the hypercube satisfies the merging condition if it does, then all the points belonging to that particular hypercube are guaranteed to satisfy the merging condition as well.


\textbf{Choosing Representative Points:} We assign representative points to further reduce the number of comparisons needed. For $n$ dimensional data we need $(3^n-1)$ representative points. For the sake of explanation, we consider two-dimensional data. Hence for each face of the grid we define eight \textit{Representative Points.} These points are labeled as follows: Top, TopRight, Right, BottomRight, Bottom, BottomLeft, Left, and TopLeft.
Every representative point in a hypercube represents the closest point to the boundary in that hypercube. If the representative point in a particular direction is within $\epsilon$ distance from any complimentary representative point in the neighboring hypercube, then both the current hypercube and the respective neighboring hypercube can be considered as one cluster saving comparison time for other points in both boxes. For example, if the current representative point is Top point for the current hypercube, then the corresponding point for the neighboring upper point is Bottom point for that hypercube. We use a token ring approach to distribute points to these eight positions. To decide which hypercube does the point belong to, we divide the point by $\epsilon / \sqrt{2}$ in each dimension to obtain the corresponding \textit{band} in which the point lies. The intersection of these bands gives us the hypercube in which the point lies. In case the grid does not start from the origin, we perform an origin shift transformation \cite{jeffrey2004mathematics}. 
 \par  
  As the first point is entered into a new hypercube, all of these eight positions are initialized to that point. An \textit{ideal position} for representative points is defined as a case where a representative point lies on the edge of the hypercube. Now whenever a new point is encountered, each position calculates the distance between the ideal position and the current point and the point being examined. If a new point is found to have a smaller distance, the corresponding representative point is updated with a new point. Multiplicity is allowed; that is one point that can represent multiple positions. This process is repeated for all the points belonging to a particular hypercube.

\begin{figure}[bpht!]
\centering
\includegraphics[scale=0.55]{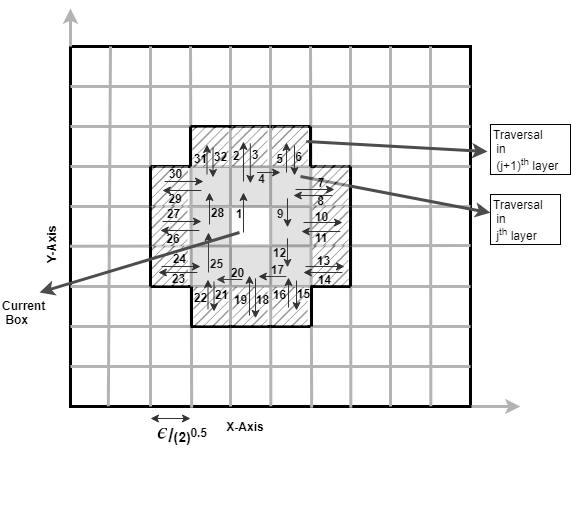}
\vspace{-1cm}
\caption{This figure shows the concept of layering that is employed by the algorithm. The central highlighted box is the box under consideration, its immediate neighbors constitute the $j^{th}$ layer and the second-degree neighbors constitute the $j+1^{th}$ layer. The arrow marks show the depth-first traversal that HCA-DBSCAN executes to find boxes that satisfy the merging condition.  }
\label{fig:grid}
\end{figure}

\textbf{Depth First Search: } We traverse the entire grid in depth first fashion. For each individual dimension d, we need to consider $(2\ceil{\sqrt{d}}+1)$ neighbouring hypercubes. Thus, when we combine all the dimensions together, the total number of neighbor hypercubes that the algorithm needs to consider is $((2\ceil{\sqrt{d}}+1)^d)-(C+1)$  for every given current box, where C+1 is a constant indicating the number of corner hypercubes (C) and the current hypercube itself which are not included in the set of neighbor boxes. Please refer \cite{boonchoo2018efficient} for in-depth explanation. For this example, the neighbor hypercubes for the current hypercube are those 20 hypercubes (shaded hypercubes $+$ line-shaded boxes) as shown in Fig. \ref{fig:grid}. We start with the hypercube closest to the shifted origin. At any given point of time, we compare the two closest representative points between hypercubes. For example, in a two-dimensional space, we would begin by checking if the \textit{Top} representative point of the current hypercube $i$ is within a $\epsilon$ - distance of the \textit{Bottom} representative point of the top neighbor hypercube $j$. We do this traversal in a clockwise fashion. If the distance between the corresponding points is less than $\epsilon$, the merging condition is said to be satisfied, and all the points in both these hypercubes are labeled to belong to the same cluster. If the merging condition fails, the next hypercube in the DFS traversal is checked, and the cluster IDs are not updated. 


\textbf{Layering:}
Two points belong to the same cluster if the distance between them is less than $\epsilon$. We do not know the location of the points inside the boxes. Therefore it is entirely possible that the distance between points in two consecutive boxes can be less than $\epsilon$. That is if we consider only neighboring boxes in our depth-first traversal, the \textit{j-layer}, we are bound to miss points that are at a distance which is less than $\epsilon$.
Consequently, the accuracy of the algorithm suffers. To overcome this problem, we introduce a concept of \textit{layering}. If the traversal for the $j$-layer box returns a failure, we check for the $(j+1)^{th}$ layer hypercube in the same direction, subject to the condition that the distance between the representative points is less than $\epsilon$. We have to check the $(j+1)^{th}$ layer box only for non-diagonal neighbors which again reduces the number of iterations. We can do this because the grids are designed in such a way that the diagonal of each hypercube is equal to the value of $\epsilon$; therefore two points in the diagonal direction cannot be at a distance less than $\epsilon$ and not lie in consecutive boxes.

\section{Experiments and Results}
\label{sec:d}
To check the efficacy and efficiency of the proposed HCA-DBSCAN algorithm, we conduct multiple experiments. We run the DBSCAN algorithm and the FastDBSCAN algorithm \cite{fastdbscan} on the same data sets with the same input parameters and compare the accuracy of the results as well as the time taken by each algorithm to achieve those results. We run the algorithm on five different data sets obtained from the UCI Machine Learning Repository \cite{ucirepo}, the data is treated in a similar fashion as \cite{gan2015dbscan}, and \cite{reiss2012introducing}. Experiments are carried out on a system running Windows 10 operating system with four Intel Processor Xeon LGA2011$-$3 E5$-$2609V3 and 128 GB RAM. The details of each data set used of the experimental analysis are given in Table \ref{table:data}. The run time obtained for each algorithm is summarized in Table \ref{table:results}. It also captures the percentage performance improvement (PPI) which is the relative improvement in runtime to the original DBSCAN algorithm.

\begin{table}[bpht!]
\centering
\caption{Details of the data sets}
\begin{tabular}{ |c|c|c| } 
 \hline
 Data set & Number of Objects & Dimension \\ \hline
 
Vicon Action data set (Case 1) & 5045 & 27 \\ 
Vicon Action data set (Case 2) & 3853 & 54 \\

PAMAP2 & 3,850,505 & 54 \\ 

Household & 2,075,259 & 7 \\

Leaf data set & 340 & 16 \\
\hline

\end{tabular}
\label{table:data}
\end{table}

\begin{table*}[bpht!]
\centering
\caption{Execution Time For Each Algorithm (in minutes)}
%
%
%

 \begin{tabular}{|c|c|c|c|c|c|}
    \hline
     & DBSCAN & \multicolumn{2}{c|}{FastDBSCAN} & \multicolumn{2}{c|}{HCA-DBSCAN}\\
    \cline{3-6}
     Data set& \makecell{with \\Spatial \\Indexing\cite{dbscan}}&Runtime & PPI & Runtime & PPI \\
    \hline
    Vicon Action Data Set (Case 1) & 2.751 & 2.384 & 13.3\% & 1.581 & 42.5\% \\ 

    Vicon Action Data Set (Case 2) & 2.5 & 1.50 & 39.8\% & 1.24 & 50.4\%\\ 

    PAMAP2 & 17,364.89 & 15,385.39 & 11.41\% & 11,704.01 & 32.6\% \\

    Household & 1709 & 1534.68 & 10.2\% & 1189.46 & 30.4\% \\

    Leaf Data Set & 0.66 & 0.48 & 26.1\% & 0.27 & 58.27\%\\
    \hline
\end{tabular}
\label{table:results}
\end{table*}



From Table \ref{table:results}, we see that the proposed algorithm achieves a significant speedup. It shows that the algorithm slows down for higher dimensions due to the increase in the number of the neighbor box as the dimensions increases. However, the algorithm retains its versatility vis-a-vis the original DBSCAN algorithm as it is accurately able to identify clusters with an arbitrary shape for every data set. 
\section{Conclusion}
\label{sec:f}
 DBSCAN has found numerous applications across domains. The main factors which have contributed to this widespread popularity are the algorithm's ability to identify arbitrarily shaped clusters, no dependence on the user for the number of clusters, and not being sensitive to outliers and noise. Traditional DBSCAN algorithm has a time complexity of $O(n^2)$ which reduces to $O(n\log{n})$ when spatial indexing like R-tree or X-tree is used. However, the complexity again rises to $O(n^2)$ for high dimensional data. We propose a new algorithm which runs in $O(n\log{n})$ for lower dimensions and scales to $O(n^{3/2})$ for higher dimensions. This algorithm uses a grid-based approach to overlay a grid on top of the data set such that all points within a box are within a $\epsilon$ - distance of each other. Therefore if one of the points in the box belongs to a particular cluster all the other points in that box will belong to that cluster as well. We use this key insight to gain a significant computational speedup over other improvements of DBSCAN. We define representative points and then traverse the grid in depth first fashion by including a concept of layering. Doing this reduces the number of computations drastically. The experimental results confirmed that the proposed algorithm has a clear advantage over original DBSCAN and Fast DBSCAN in terms of computational time while preserving accuracy.

\bibliographystyle{IEEEtran}
\bibliography{NeurIPS}

\begin{thebibliography}{10}
\providecommand{\url}[1]{#1}
\csname url@samestyle\endcsname
\providecommand{\newblock}{\relax}
\providecommand{\bibinfo}[2]{#2}
\providecommand{\BIBentrySTDinterwordspacing}{\spaceskip=0pt\relax}
\providecommand{\BIBentryALTinterwordstretchfactor}{4}
\providecommand{\BIBentryALTinterwordspacing}{\spaceskip=\fontdimen2\font plus
\BIBentryALTinterwordstretchfactor\fontdimen3\font minus
  \fontdimen4\font\relax}
\providecommand{\BIBforeignlanguage}[2]{{%
\expandafter\ifx\csname l@#1\endcsname\relax
\typeout{** WARNING: IEEEtran.bst: No hyphenation pattern has been}%
\typeout{** loaded for the language `#1'. Using the pattern for}%
\typeout{** the default language instead.}%
\else
\language=\csname l@#1\endcsname
\fi
#2}}
\providecommand{\BIBdecl}{\relax}
\BIBdecl

\bibitem{dbscan}
M.~Ester, H.-P. Kriegel, J.~Sander, and X.~Xu, ``A density-based algorithm for
  discovering clusters a density-based algorithm for discovering clusters in
  large spatial databases with noise,'' in \emph{Proceedings of the Second
  International Conference on Knowledge Discovery and Data Mining}, ser.
  KDD'96, 1996, pp. 226--231.

\bibitem{award}
KDD, ``2014 {SIGKDD TEST OF TIME AWARD},''
  \url{http://www.kdd.org/News/view/2014-sigkdd-test-of-time-award}, 2014,
  [Online; accessed 10-October-2016].

\bibitem{wiki:db}
Wikipedia, ``{DBSCAN}--- wikipedia{,} the free encyclopedia,''
  \url{https://en.wikipedia.org/wiki/DBSCAN}, 2016, [Online; accessed
  10-October-2016].

\bibitem{rtree}
S.~Berchtold, D.~Keim, and H.~Kriegel, ``The {X-tree}: An efficient and robust
  access method for points and rectangles,'' in \emph{Proc. 1996 Int. Conf.
  Very Large Data Bases}, 1996, pp. 28--39.

\bibitem{wiki:hypercube}
Wikipedia, ``Hypercube--- wikipedia{,} the free encyclopedia,''
  \url{https://en.wikipedia.org/wiki/HypercubeN}, 2016, [Online; accessed
  20-October-2016].

\bibitem{jeffrey2004mathematics}
A.~Jeffrey, \emph{Mathematics for engineers and scientists}.\hskip 1em plus
  0.5em minus 0.4em\relax CRC Press, 2004.

\bibitem{boonchoo2018efficient}
T.~Boonchoo, X.~Ao, and Q.~He, ``{An Efficient Density-based Clustering
  Algorithm for Higher-Dimensional Data},'' \emph{arXiv preprint
  arXiv:1801.06965}, 2018.

\bibitem{fastdbscan}
S.~J. Nanda and G.~Panda, ``Design of computationally efficient density-based
  clustering algorithms,'' \emph{Data \& Knowledge Engineering}, vol.~95, pp.
  23--38, 2015.

\bibitem{ucirepo}
\BIBentryALTinterwordspacing
D.~Dheeru and E.~Karra~Taniskidou, ``{UCI} machine learning repository,'' 2017.
  [Online]. Available: \url{http://archive.ics.uci.edu/ml}
\BIBentrySTDinterwordspacing

\bibitem{gan2015dbscan}
J.~Gan and Y.~Tao, ``{DBSCAN} revisited: mis-claim, un-fixability, and
  approximation,'' in \emph{Proceedings of the 2015 ACM SIGMOD International
  Conference on Management of Data}.\hskip 1em plus 0.5em minus 0.4em\relax
  ACM, 2015, pp. 519--530.

\bibitem{reiss2012introducing}
A.~Reiss and D.~Stricker, ``Introducing a new benchmarked dataset for activity
  monitoring,'' in \emph{Wearable Computers (ISWC), 2012 16th International
  Symposium on}.\hskip 1em plus 0.5em minus 0.4em\relax IEEE, 2012, pp.
  108--109.

\bibitem{adefaster}
A.~Ade~Gunawan, ``A faster algorithm for {DBSCAN},'' \emph{MSc Thesis, TU
  Eindhoven}, 2013.

\end{thebibliography}
\newpage
\section{Supplementary Material}

\begin{figure}[bpht!]
\centering
\includegraphics[width=8cm, height=14cm]{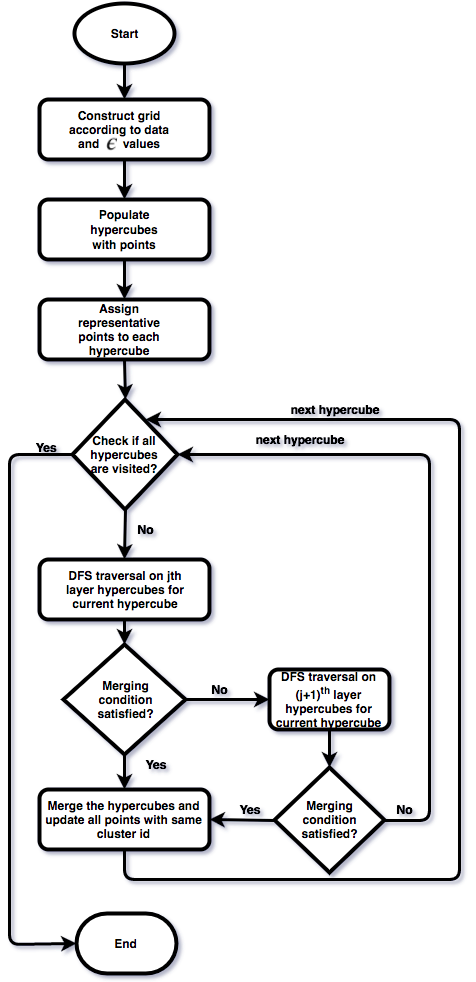}
\caption{Flowchart detailing the steps followed in the proposed HyperCube based Accelerated DBSCAN}
\label{fig:HCA-DBSCAN}
\end{figure}

The steps followed in the proposed HyperCube Accelerated DBSCAN algorithm are shown in Fig. \ref{fig:HCA-DBSCAN}.

\section{Complexity Analysis}
For HCA-DBSCAN, we have to go through all the non-empty hypercubes and mark them as non-empty which takes $O(n)$ time where n number of non-empty boxes. Also, considering lower dimensions, checking whether a box is to be merged with other neighbouring box by finding the nearest representative point from a neighboring box’s representative points, requires $O(\log(n))$ time \cite{adefaster}. Thus, the total time complexity is $O(n\log(n))$ . For higher dimension, this algorithm takes $O(n.T_d)$ where $T_d$ is the time taken for cluster merging in d dimension. If time taken for iterating in every dimension is $(2\ceil{\sqrt{n}}+1)$, then we get total time complexity of $O(n.(2\ceil{\sqrt{n}}+1))$ which is further simplified to $O(n^(3/2))$ \cite{boonchoo2018efficient}. 

\section{The DBSCAN Algorithm}
\label{sec:a}
In this section, we explain the original DBSCAN algorithm \cite{dbscan}. The algorithm takes two parameters as inputs $\epsilon$ and \textit{MINPTS}. $\epsilon$ governs the density distribution of the data and MINPTS gives information about the minimum number of points that are required to be present within the $\epsilon $-radius of a given point for it to be considered \textit{'dense'}. To understand the algorithm we need to define the following terms.

\begin{definition}[$\epsilon $-neighborhood of a point]
The $\epsilon $ - neighborhood of a point $p$, denoted by $N_{\epsilon }(p)$, is defined by
$N_\epsilon(p)=\{q\in D|dist(p,q)\leq\epsilon\}$, where $dist()$ is the distance calculate operator between $p$ and $q$, and $D$ is the set of all points in the data set. 
\end{definition}

\begin{definition}[Directly density-reachable]
A point $p$ is directly density-reachable from a point $q$ wrt. $\epsilon$, and  MINPTS if
\begin{enumerate}
    \item $p \in N_{\epsilon }(q)$ i.e., $p$ belongs to the $\epsilon $-neighborhood of $q$ and
    \item $|N_{\epsilon }(q)| \geq$ MINPTS (core point condition).
\end{enumerate}
\end{definition}

\begin{definition}[Density-reachable]
A point $p$ is \textit{density-reachable} from a point $q$ wrt. $\epsilon$, and MINPTS if there is a chain of points $p_1,p_2,\ldots,p_n$, $p_1=q$ and $p_n=p$ such that $p_{i+1}$ is directly density-reachable from $p_i$.
\end{definition}

\begin{definition}[Density-connected]
A point $p$ is \textit{density connected} to a point $q$ wrt. $\epsilon$, and MINPTS if there is a point $o$ such that both, $p$ and $q$ are density-reachable from $o$ wrt. $\epsilon$, and MINPTS.
\end{definition} 

\begin{definition}[Cluster]
A cluster $C$ wrt. $\epsilon$, and MINPTS is a non-empty subset of the data set $D$ satisfying the following conditions:
\begin{enumerate}
    \item $\forall p, q$: if $p\in C$ and $q$ is density-reachable from $p$ wrt. $\epsilon$, and MINPTS, then $q\in C$. (Maximality) 
    \item $\forall p,q \in C$: if $p$ is density-connected to $q$ wrt. $\epsilon$, and MINPTS. (Connectivity)
\end{enumerate}
\end{definition} 

\begin{figure}[bpht!]
\centering
\includegraphics[scale=0.52]{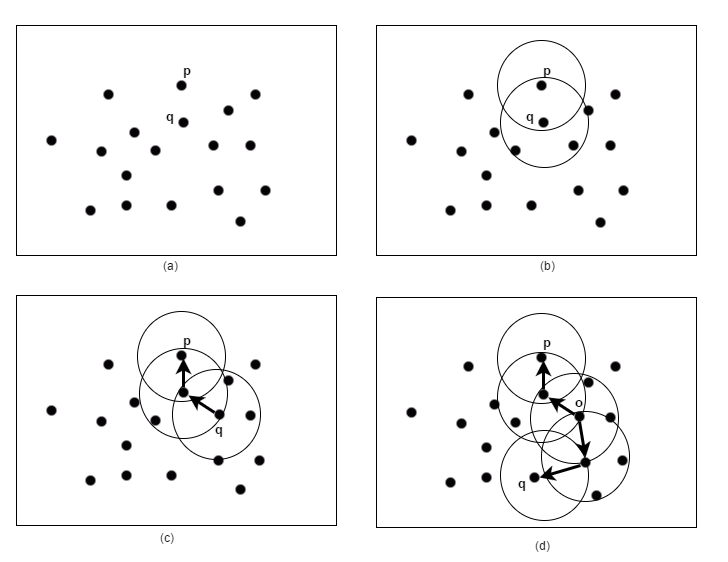}
\caption{DBSCAN definitions: (a) $p$ is a border point, $q$ is a core point (it has more than MINPTS number of points in its $\epsilon$ neighborhood) (b) $p$ is directly density reachable from $q$ but not vice-versa.(c) $p$ is density reachable from $q$ but not vice-versa (d) $p$ and $q$ are density connected to each other via point $o$.}
\label{fig:def}
\end{figure}

These definitions are visualized in Fig \ref{fig:def}. To find a cluster, DBSCAN starts with an arbitrary point $p$ and retrieves all points density-reachable from $p$ for $\epsilon $ and MINPTS. If $p$ is a core point, this procedure yields a cluster concerning $\epsilon $ and MINPTS. If $p$ is a border point (non-core point, i.e., definition 2, condition 2 is not satisfied), no points are density-reachable from $p$, and DBSCAN visits the next point of the data set. Thus in the worst case, when spatial indexing is not used to speed up the nearest neighbor search, each point is compared to every other point in the data set, yielding a time complexity of $O(n^2)$.

\section{Algorithms}
\begin{algorithm*}[bpht!]
\SetAlgoLined
\SetKwInOut{Input}{input}
\SetKwInOut{Output}{output}
\SetKwInOut{Function}{function}
\Function{GENERATE-HYPERCUBE(data set D, $\epsilon$)}
\Input{data set $D$ containing all input points and $\epsilon$ user input parameter}
\Output{HyperCubeDetailsNbPos : dictionary with hypercube coordinates as key and its updated neighbouring, position details as value}

\ForEach{dimension $i$ in n}{
// n is number of dimension of data set D which is globally defined\\
\emph{$i_{\text{Band}}$ $\leftarrow$ generate steps of $\frac{\epsilon}{\sqrt D}$ by incrementally increasing from shifted origin to boundary of the hypercube container for every axis   \;}
}

\ForEach{StepValue $v$ in every $i_{\text{Band}}$}{

HyperCubeList $\leftarrow$ generate all the combination of hypercube for every axis using StepValue $v$ \;
}
\ForEach{HyperCubePoint $k$ in HyperCubeList}{
Position $\leftarrow$ IDENTIFY-POSITION-OF-HYPERCUBE(k)\;
HyperCubeDetailsNbPos $\leftarrow$ APPEND( $k$ , NEIGHBOURING-POINTS(k), Position)\;
}

\Return{HyperCubeDetailsNbPos}\;
\caption{HyperCube creation and initialization}
\end{algorithm*}
\vspace{-1mm}

\begin{algorithm*}[bpht!]
\SetAlgoLined

\SetKwInOut{Input}{input}
\SetKwInOut{Output}{output}
\SetKwInOut{Function}{function}
\Function{COMPUTE-REPRESENTATIVE-POINTS(data set D, $\epsilon$)}
\Input{data set $D$ containing all input points and $\epsilon$ user input parameter }
\Output{HyperCubeDetailsRepPts : dictionary of hypercube coordinate as key and its updated representative points as value}

\ForEach{point  $j$ in data set $D$ }{

IdentifiedHyperCube $\leftarrow$ FIND-CORRESPONDING-HYPERCUBE-OF-POINT($j$)\;
Status $\leftarrow$ CHECK-VISITED(IdentifiedHyperCube)

\If(\tcp*[h]{visiting for the first time}){Status is not visited}
{

mark IdentifiedHyperCube as visited\;
initialize all the $(3^n)-1$ representative points to the current $j$ point\tcp*{n is number of dimension for data set D which is globally defined}
}
\Else{
\ForEach{representative point $l \in $ set($(3^n)-1$ representative points)}{
compare distance between the ideal point-and-the current point represented by $a$ and ideal point-and-representative point $l$ represented by  $b$ \tcp*{ideal points are ideal representative points on the edge of hypercube }
\If{$a<b$}{
representative point $l$ $\leftarrow$ current point $j$ \;}
}
}
HyperCubeDetailsRepPts $\leftarrow$ insert the current point $j$ in IdentifiedHyperCube with its updated $(3^n)-1$ representative points \;
}
\Return{HyperCubeDetailsRepPts}\;

\caption{Computation of Representative Points }
\end{algorithm*}

\vspace{-1cm}
\begin{algorithm*}[tbh!]
\SetAlgoLined
\newcommand\mycommfont[1]{\footnotesize\ttfamily\textcolor{black}{#1}}
\SetCommentSty{mycommfont}
\SetKwInOut{Input}{input}
\SetKwInOut{Output}{output}
\SetKwInOut{Function}{function}
\Function{CLUSTERING-FUNCTION( HyperCubeDetails, CurrentHyperCube, $\epsilon$)}
\Input{CurrentHyperCube : hypercube under consideration, HyperCubeDetails : dictionary with hypercube coordinate as key and its updated neighbouring points, representative points, position details as value and $\epsilon$ : user input parameter indicating the density of the clusters needed.}
\Output{ClusterList = list of total number of cluster and points in each cluster}

\If{CurrentHyperCube is not visited}{
\If{CurrentHyperCube is opened}{
\ForEach{NeighbouringHyperCube of CurrentHyperCube }{
\If{distance between corresponding representative points $< \epsilon $ for NeighbouringHyperCube in $j^{th}$ layer }{
ClusterID$_{\text{NeighbourHyperCube}}$ $\leftarrow$
ClusterID$_{\text{CurrentHyperCube}}$\tcp*{merging cluster}
mark CurrentHyperCube as visited\;
// recursive call\\
CLUSTERING-FUNCTION(NeighbourHyperCube, HyperCubeDetails, $\epsilon$)\;
}
\Else{
//check for NeighbouringHyperCubes which are not diagonal hypercube as distance greater than $\epsilon$ for diagonal hypercubes\\
\If{the NeighbouringHyperCube is not a diagonal hypercube}{
\If{distance between corresponding representative points $< \epsilon $ for NeighbouringHyperCube in $(j+1)^{th}$ layer hypercube }{
ClusterID$_{\text{NeighbourHyperCube}}$ $\leftarrow$ ClusterID$_{\text{CurrentHyperCube}}$\tcp*{merging cluster}
mark CurrentHyperCube as visited\;
// recursive call\\
CLUSTERING-FUNCTION(NeighbourHyperCube, HyperCubeDetails, $\epsilon$)\;
}
}
}
}
}
}
\Return{ClusterList}\;
\caption{Clustering mechanism for HyperCube}
\end{algorithm*}

\begin{algorithm*}[tbh!]
\SetAlgoLined
\newcommand\mycommfont[1]{\footnotesize\ttfamily\textcolor{black}{#1}}
\SetCommentSty{mycommfont}
\SetKwInOut{Input}{input}
\SetKwInOut{Output}{output}
\SetKwInOut{Function}{function}
\Function{HYPERCUBE-BASED-ACCELERATED-DBSCAN(data set $D$, $\epsilon$)}
\Input{data set $D$ containing all input points, $\epsilon$ user input parameter indicating the density of the clusters needed.}
\Output{FinalClusterList = aggregation of individual ClusterList for each hypercube}

initialize CurrentClusterID=1\;
HyperCubeDetails1 $\leftarrow$ GENERATE-HYPERCUBE(data set D, $\epsilon$)\;
HyperCubeDetails2 $\leftarrow$ COMPUTE-REPRESENTATIVE-POINTS(data set D, $\epsilon$)\;
HyperCubeDetails $\leftarrow$ HyperCubeDetails1 + HyperCubeDetails2

\ForEach{hypercube  $k$ in HyperCubeDetails }{

\If{ $k$ is not visited and $k$ has some points}{
mark $k$ as visited\;
ClusterID$_{k}$ $\leftarrow$ CurrentClusterID\;
FinalClusterList $\leftarrow$ FinalClusterList + CLUSTERING-FUNCTION(HyperCubeDetails, current hypercube $k$, $\epsilon$)

CurrentClusterID = CurrentClusterID+1\;
}
\Else{pass to the next hypercube in HyperCubeDetails
}
}
\Return{FinalClusterList}\;
\caption{HyperCube based Accelerated DBSCAN}
\end{algorithm*}


\




\end{document}